\documentclass{aa}  

\usepackage{graphicx}
\usepackage{txfonts}
\usepackage{color}
\usepackage{lipsum}
\usepackage{subcaption}         
\usepackage{lscape}
\usepackage{placeins}
\usepackage[normalem]{ulem}
\usepackage{hyperref}

\newcommand{\feh}{{[Fe/H]}\xspace}
\newcommand{\gaia}{{\it Gaia}\xspace}
\newcommand{\kms}{{$\rm{km}\,\rm{s}^{-1}$}\xspace}
\newcommand{\kpckms}{{$\rm{kpc\, km}\,\rm{s}^{-1}$}\xspace}

\begin{document}

   \title{Searching for stars ejected from the Galactic Centre in DESI}

   \author{Sill Verberne\inst{1}\thanks{verberne@strw.leidenuniv.nl} \and Sergey E.~Koposov\inst{2,3} \and Elena Maria Rossi\inst{1} \and Zephyr Penoyre\inst{1}
        }

      \institute{Leiden Observatory, Leiden University,
              P.O. Box 9513, 2300 RA Leiden, the Netherlands
         \and
             Institute for Astronomy, University of Edinburgh, Royal Observatory, Blackford Hill, Edinburgh EH9 3HJ, UK
         \and
             Institute of Astronomy, University of Cambridge, Madingley Road, Cambridge CB3 0HA, UK
             }
   \date{Received XXX / Accepted YYY}

  \abstract
  {Dynamical interactions between stars and the supermassive black hole Sgr A* at the Galactic Centre (GC) may result in stars being ejected into the Galactic halo. While recent fast ejections by Sgr A* have been identified in the form of hypervelocity stars (hundreds to thousands of km/s), it is also believed that the stellar halo contains slower stars, ejected over the last few billion years. In this study we used the first data release of DESI to search for these slower GC ejecta, which are expected to stand out from the stellar halo population thanks to their combined high metallicity (${\rm [Fe/H]}\gtrsim0$) and low vertical angular momentum ($L_Z$), whose distribution should peak at zero. Our search did not yield a detection but allowed us to place an upper limit on the ejection rate of stars from the GC of $\sim2.8\times10^{-3}$ yr$^{-1}$ over the past $\sim5$ Gyr, which is ejection model independent. This implies that our result can be used to put constraints on different ejection models, including those that invoke mergers of Sgr A* with other massive black holes in the last last few billion years.}

   \maketitle

\section{Introduction}
The stellar halo is a highly complex environment made up of substructures containing information on the assembly history of the Galaxy. In particular, with the advent of \gaia \citep{Gaia_2016}, a wealth of knowledge has been gathered on the individual components of the halo \citep[for a recent review, see][]{Deason_2024}. We now understand that the halo is perhaps entirely made up of the debris of past merger events. This means that the halo provides a means of studying the remains of high-redshift dwarf galaxies in great detail given the relative proximity of halo stars. 

In our understanding of the halo, we rarely consider the effect of the Galactic Centre (GC) and in particular the supermassive black hole Sagittarius A* (Sgr A*). However, there is reason to think that this might not be entirely justified. In recent decades, it has become clear that Sgr A* ejects stars on unbound trajectories travelling through the halo via the Hills mechanism \citep{Hills_1988, Yu_2003, Brown_2005, Brown_2015, Koposov_2020}. Moreover, a significant fraction of these ejected stars are expected to remain bound to the Galaxy \citep[e.g.][]{Bromley_2006, Rossi_2014}. This means that there would be a population of stars with the chemical properties of the GC that populate the halo \citep{Brown_2007}. Stars ejected from the GC are often referred to as hypervelocity stars, but here we refer to them as GC ejecta since we focus on the stars still bound to the Galaxy, rather than the unbound population.
 
Galactic Centre ejecta provide interesting science cases since they are windows into the complex GC environment in parts of the sky that are easier to study, while unlocking wavelength ranges for observations that are inaccessible for sources at the GC. In addition, their trajectories can be used as tracers of the Galactic potential \citep[e.g.][]{Contigiani_2019, Gallo_2022, Armstrong_2025}. However, their application has been limited so far because only a single unambiguous star ejected from the GC is known \citep{Koposov_2020}, alongside a handful of promising candidates \citep{Brown_2005, Brown_2014, Brown_2018}. A big factor in the difficulty to accurately identify GC ejecta is that many of these candidates tend to be distant, which makes their past trajectories uncertain. The rate and properties of GC ejecta are therefore also still uncertain \citep{Zhang_2013, Brown_2014, Marchetti_2022, Evans_2022_I, Evans_2022}. The best constraints on the rate so far have been published in \citet{Verberne_2024, Verberne_2025}, who find an upper limit of $10^{-5}$ yr$^{-1}$ for stars more massive than 1 M$_{\odot}$ and a higher ejection rate over the past 10 Myr of about $10^{-4}$ yr$^{-1}$.

Additionally, past mergers of Sgr A* with intermediate-mass or supermassive black holes would have boosted the rate of stars ejected from the GC by orders of magnitude \citep[e.g.][]{Yu_2003, Gualandris_2005, Baumgardt_2006, Levin_2006, Evans_2023}. Furthermore, kicks from stellar-mass black holes and the disruption of infalling dwarf galaxies might have also produced GC ejecta \citep{O'Leary_2008, Abadi_2009}. For these reasons, it is unknown how large the halo population of GC ejecta is.

In this work we aim to constrain the population of GC ejecta that has built up in the stellar halo over the lifetime of the Galaxy. We expect that two characteristic features can be used to identify these stars: they should be metal-rich since they originated in the GC \citep[e.g.][]{Schultheis_2019, Schodel_2020}, and their initial orbital angular momentum should be zero since they are ejected radially from the GC.

No other stellar halo populations with metallicities similar to that of the GC are known, making metallicity a crucial factor in the identification of this population of stars. We used the recently released Dark Energy Survey Instrument \citep[DESI;][]{DESI_2016} Data Release 1 \citep[DR1;][]{DESI_dr1} in combination with \gaia DR3 \citep{Gaia_2023} to calculate the angular momenta of individual stars. We used the distributions of angular momenta and \feh to search for a statistical overdensity of low-angular-momentum, high-metallicity stars.

The structure of this work is as follows: We start by discussing the origin of GC ejecta and our simulations in Sect.~\ref{sec:GC_ejecta}, which is followed by a description of the observations we used in Sect.~\ref{sec:obs}. In Sect.~\ref{sec:id} we discuss our method for identifying a possible population of GC ejecta before presenting our results in Sect.~\ref{sec:results}. In Sect.~\ref{sec:discussion} we discuss our assumptions and put our results into perspective. Finally, in Sect.~\ref{sec:summary} we summarise our work.

\section{Galactic Centre ejecta}\label{sec:GC_ejecta}
In this section we start by discussing a mechanism through which stars can be ejected from the GC and end up in the stellar halo, before we describe how we implemented this in our simulations. We explain the specific ejection mechanism used in our simulations below, but we argue that our results are in fact ejection model independent (see Sect.~\ref{sec:ej_model}).

\subsection{Hills mechanism}
The Hills mechanism \citep{Hills_1988} acts on a stellar binary that approaches a massive black hole within the tidal radius, where the tidal force of the black hole separates the stellar binary, ejecting one star and capturing its companion \citep[for a review, see][]{Brown_2015}. The ejected stars can gain enough energy to become unbound to the Galactic potential. For a contact binary progenitor, for instance, the ejected star can be ejected at up to about 3500 \kms \citep{Rossi_2014}. However, a significant fraction of ejected stars will remain bound to the Galaxy, travelling on highly eccentric orbits. Although the fraction of stars that remain bound to the Galaxy is a function of the binary progenitor population properties, in particular the mass ratio and semi-major axis distributions, a population of stars ejected from the GC will accumulate in the stellar halo. For a star ejected from the GC on a bound orbit in a perfect axisymmetric potential, we expect $L_Z$ to be conserved and equal to 0, while $L_X$ and $L_Y$ will oscillate due to the gravity of the disc.

The number of stars in this population is highly uncertain mainly because the ejection rate is only constrained to an order of magnitude and the progenitor binary properties are poorly constrained. Most literature estimates of the ejection rate are between $10^{-5}$ and $10^{-3}$ yr$^{-1}$ \citep{Hills_1988, Yu_2003, Zhang_2013, Brown_2014, Marchetti_2022, Evans_2022, Verberne_2024} and are based on recent (past $\sim100$ Myr) ejections. \citet{Verberne_2025} extend this with a lower limit of the ejection rate over about a billion years of $\sim10^{-5}$ yr$^{-1}$.

\subsection{Simulations}\label{sec:simulations}
To predict the physical and observational footprint of the population of stars ejected from the GC, we performed a suite of simulations. Our starting point was the simulations presented in \citet[][their Sects. 3 and 4]{Verberne_2025}, which we refer to for a detailed description. To summarise, we sampled from a progenitor binary population defined by the star formation history, initial mass function, log-period distribution, and mass ratio distribution. In \citet{Verberne_2025} two progenitor populations are described: a young population with an average ejection rate over the past $\sim10$ Myr of approximately $10^{-4}$ yr$^{-1}$ and an old population with an ejection rate of at least $\sim10^{-5}$ yr$^{-1}$ over potentially billion-year timescales. We focussed here on the old population but note that past bursts of star formation near Sgr A* might have boosted the ejection rate, similar to what seems to have happened recently with the formation of the clockwise disc. 

We randomly selected one of the stars in the progenitor binary to be ejected regardless of its mass, as appropriate for incoming centre of mass parabolic trajectories, \citep[which is realistic;][]{Sari_2009, Kobayashi_2012}. We assumed that stars are isotropically ejected from the GC and evolved their orbits using {\tt Agama} \citep{Agama_2019} in the \citet{McMillan_2017} Galactic potential until the `present time'. We considered stars ejected over the past 5 Gyr to allow significant accumulation of ejected stars in the halo. We evaluated our choice of potential by comparing it with a non-axisymmetric potential in Sect.~\ref{sec:potential}.

To obtain observational parameters for our simulated stars, in particular the star brightness in different photometric bands, we utilised {\tt MIST} isochrones \citep{Mist0, MistI}. This allowed us to forward-model the population of ejected stars observable by DESI \citep{DESI_dr1}. Furthermore, we used {\tt PyGaia}\footnote{\url{https://github.com/agabrown/PyGaia}} to obtain estimated uncertainties on the measured parallax of stars in our simulations.

\subsection{Results from simulations}\label{sec:sim_results}
\begin{figure}
    \centering
    \includegraphics[width=\linewidth]{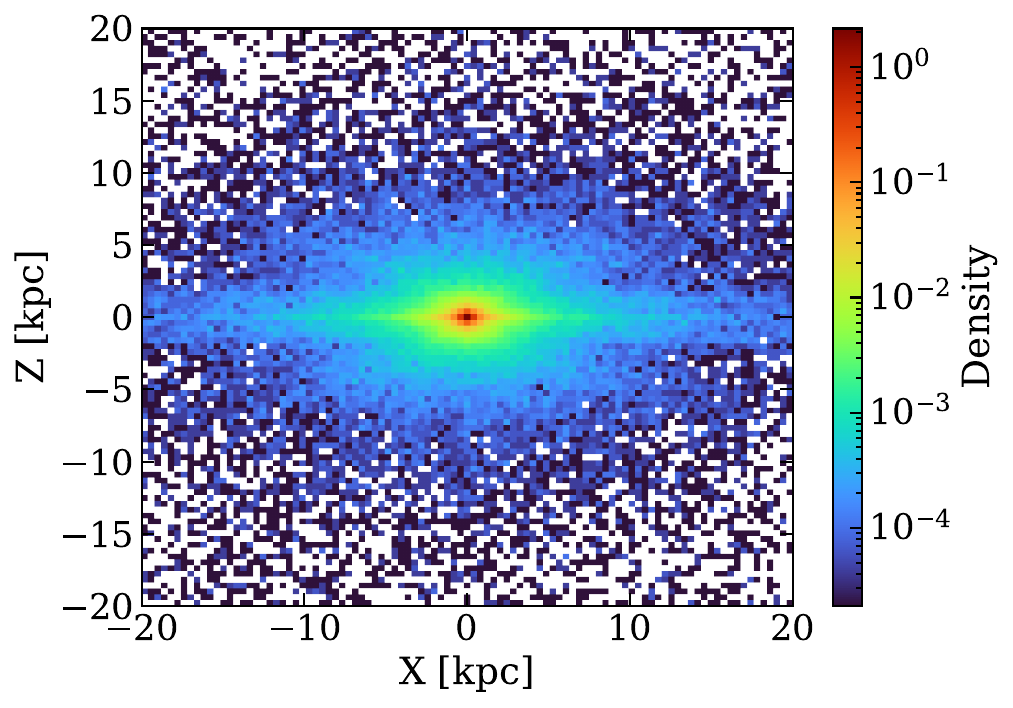}
    \caption{Galactic Cartesian X--Z distribution of the GC ejecta propagated in the \citet{McMillan_2017} potential over 5 Gyr.}
    \label{fig:XZ_sim}
\end{figure}

\begin{figure}
    \centering
    \includegraphics[width=\linewidth]{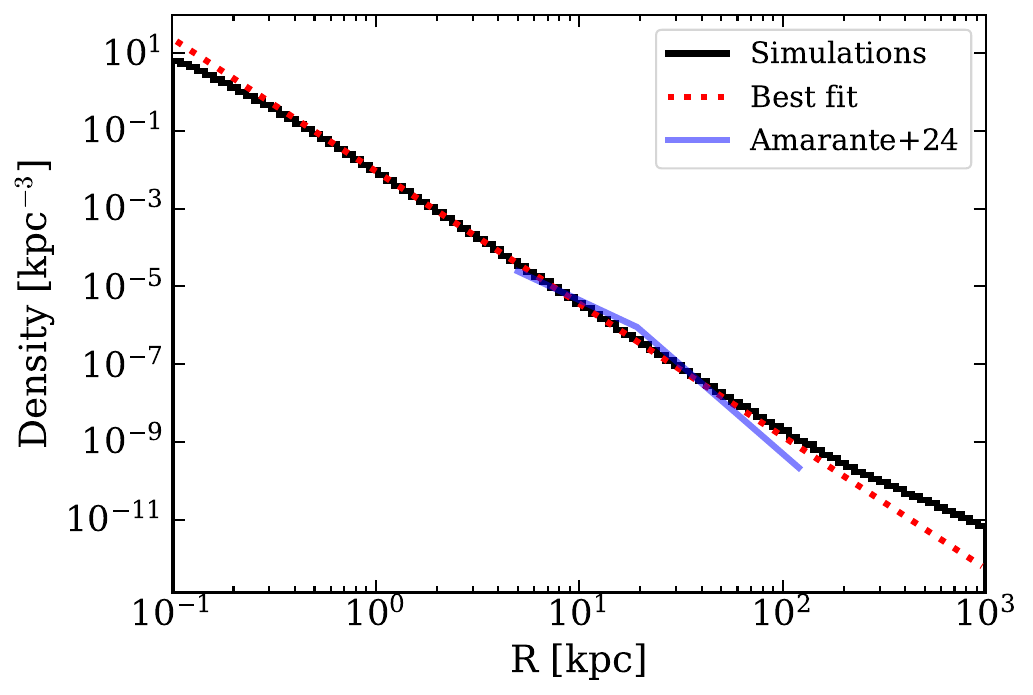}
    \caption{Radial density profile of the population of stars ejected from the GC in the \citet{McMillan_2017} potential. The red line shows the best fit power law in the range [1, 100] kpc, $\rho\propto r^{-3.41\pm0.01}$. For reference, we show the halo profile found in \citet{Amarante_2024}, which we scaled to our data at the solar circle.}
    \label{fig:radial_density}
\end{figure}

Now that we have presented our simulations, we will discuss some of the key results they provide, which will help guide our search for GC ejecta in DESI DR1. We only analysed ejected stars with an apocentre above 0.1 kpc for computational reasons and because we are mainly interested in the ejected stars that reach the stellar halo. Firstly, we examined the distribution of GC ejecta in Galactocentric Cartesian coordinates (see Fig.~\ref{fig:XZ_sim}).
Although stars are isotropically ejected from the GC, we can see that gravitational focussing by the disc causes many ejected stars to end up in or near the stellar disc, which is in the $Z=0$ plane \citep[see also][]{Kenyon_2018}. Of particular interest compared to stellar halo populations is the radial density profile, which we show in Fig.~\ref{fig:radial_density}.
In our definition, the fraction of stars $n$ is calculated from the density $\rho(r)$ as $n = \int4\pi r^2\rho(r)\ {\rm d}r$. We performed a least-squares fit using a power law of the form $\rho(r) = \rho_0 r^{-\alpha}$ to the range [1, 100] kpc. We find a power-law slope of $\alpha = 3.41\pm0.01$, which is significantly steeper compared to the stellar halo within $\sim20$ kpc but tends to be less steep compared to measurements of the stellar halo outside $\sim20$ kpc \citep[e.g.][]{Bell_2008, Xue_2015, Amarante_2024, Yu_2024}. Moreover, at distances above about 100 kpc there is an overdensity compared to the simple power-law slope caused by unbound stars. The number and velocity distribution of unbound GC ejecta depends on the ejection model. \\

\begin{figure}
    \centering
    \includegraphics[width=\linewidth]{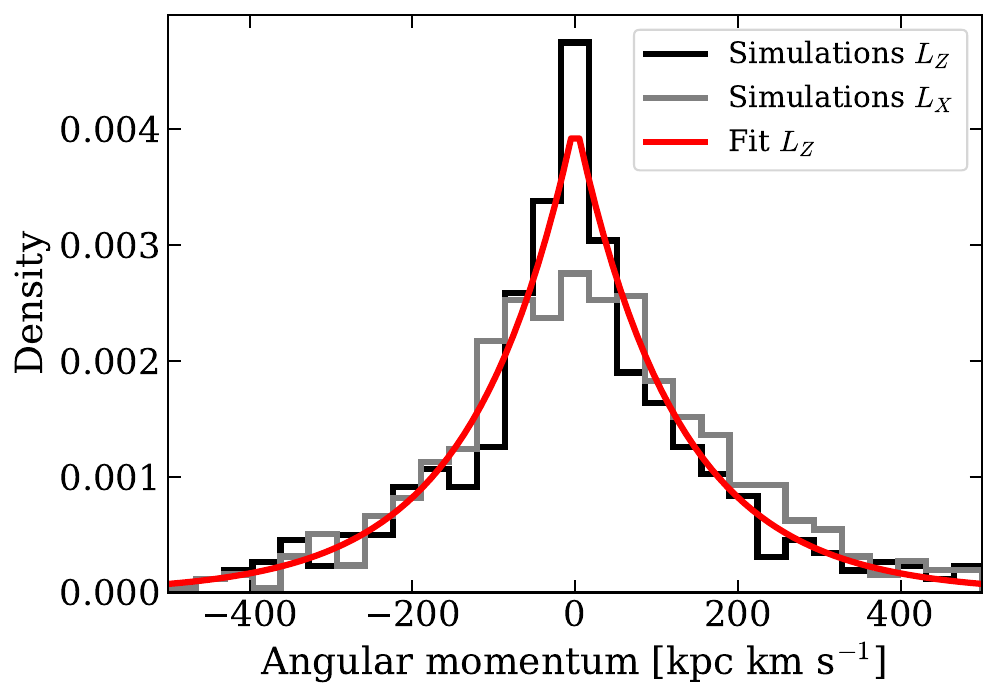}
    \caption{Measured distribution of $L_Z$ and $L_X$ for GC ejecta in the simulations. Since the true $L_Z$ of ejected stars is 0, the observed spread is due to distance uncertainties. We provide a fit to the data in red. In grey we also plot the measured $L_X$ distribution.} 
    \label{fig:Lz_sim}
\end{figure}

Another key characteristic of GC ejecta is their angular-momentum distribution. To calculate the angular momentum of a star, we needed the proper motion, radial velocity, and distance. We combined the proper motion from \gaia DR3 \citep{Gaia_2023}, the RVSpecFit (RVS) radial velocity from the DESI Milky Way Survey (MWS) value-added catalogue \citep{Koposov_2024}, and the distance from the DESI spectrophotometric distance value-added catalogue \citep{Li_2025}. However, before we can present the expected distribution of $L_Z$ in DESI, we first need to discuss the observational selection function. The one of the MWS bright survey in DESI is described in \citet{Cooper_2023} and \citet{Koposov_2025}. We approximated it as follows:
\begin{itemize}
    \item $16 < r < 19$,
    \item Dec $>-15$ deg,
    \item $|b| > 40$ deg,
    \item completeness $= 20\%$,
    \item {\tt $(g-r<0.7)\ |\ (g-r>0.7\ \wedge\ \varpi<3\sigma_\varpi+0.3)$},
\end{itemize}
where $\varpi$ and $\sigma_\varpi$ are the parallax and corresponding uncertainty, respectively. We know that the observed spread in the distribution of $L_Z$ for GC ejecta in an axisymmetric potential will be dominated by observational uncertainty in the distance to individual stars, since $L_Z=0$ intrinsically. To account for this, we randomly sampled from the distribution of distance/uncertainty for sources in the \citet{Li_2025} value-added distance catalogue, for each simulated star. We convolved the true distances to the simulated stars with these uncertainties, only considering sources with distance/uncertainty $> 5$, since these will be the sources we analyse in Sect.~\ref{sec:obs}. From our procedure, we obtained the expected observed angular-momentum distribution for GC ejecta in DESI, shown in Fig.~\ref{fig:Lz_sim}.
While the distribution in $L_Z$ is exclusively set by the distance uncertainty, the spread in $L_X$ is a combination of the intrinsic $L_X$ distribution and the distance uncertainty. We performed an unbinned fit with a Laplace distribution of the form
\begin{equation}
    f(x\ |\ \mu,b) = \frac{1}{2b}\exp\left(-\frac{|x-\mu|}{b}\right)
\end{equation}
to the $L_Z$ distribution shown in Fig.~\ref{fig:Lz_sim}, where we fix the offset $\mu = 0$. This provides us with the scale parameter for which we find the 50th percentile value $b = 1.2\times10^2$ \kpckms (narrower than other populations; see Sect.~\ref{sec:id}). We used this distribution and shape as the characteristic shape of GC ejecta in the DESI data, and evaluate the (in)dependence of the assumed ejection model on our results in Sect.~\ref{sec:ej_model}.

Given our simulations and the selection function of DESI, we can also compute the number of GC ejecta in the DESI MWS bright catalogue given an ejection rate. For GC ejecta from the old population we referenced above, we expect $\sim0.18*\frac{\eta}{10^{-5} {\rm yr^{-1}}}$ GC ejecta stars in DESI, with $\eta$ the average ejection rate over the past 5 Gyr. Furthermore, we expect that recent ejections from the young population will contribute $\sim0.27*\frac{\eta}{10^{-4}{\rm yr^{-1}}}$ GC ejecta for an ejection rate of $10^{-4}$ yr$^{-1}$ during the past 10 Myr to DESI. These estimates are likely somewhat pessimistic, since the bright programme includes some sources outside the selection function estimate that we used. Moreover, past mergers between Sgr A* and massive black holes could have additionally ejected large numbers of stars from the GC through the `slingshot mechanism' that are not accounted for in our simulations, as mentioned in the introduction.

\section{Observations}\label{sec:obs}
We have discussed the observational selection function and forward modelling of DESI. Here we describe the specific observational data from DESI used in this study.

We relied on data from DESI DR1 and in particular on the MWS catalogue \citep{Cooper_2023}. The MWS catalogue contains spectra and properties derived from $\sim5$M stars. We used the radial velocities and \feh abundance measurements from the RVS pipeline \citep{RVSpecFit_2019, Cooper_2023} in combination with the distances from the spectrophotometric MWS SpecDis catalogue \citep{Li_2025}. For the \feh measurements, we used the calibration described in \citet{Koposov_2025}, which makes the measurements accurate to $\sim0.1$ dex. Note that this recalibration is very important for our results, since high metallicity giants in DESI show a systematic offset towards too high \feh in RVS\footnote{Without the metallicity recalibration from \citet{Koposov_2025}, we find an overdensity of high-metallicity giants centred at $L_Z=0$. The recalibration shifts these stars to lower metallicities, consistent with the GSE halo population.}. 

To ensure that we only used reliable measurements, we applied the following data quality cuts in DESI:
\begin{itemize}
    \item {\tt SUCCESS} = True (RVS flag),
    \item {\tt RVS\_WARN} = 0,
    \item {\tt FEH\_ERR} $< 0.2$,
    \item {\tt RR\_SPECTYPE} = STAR,
    \item {\tt DIST/DISTERR} $>5$.
\end{itemize}

\noindent Finally, we used the crossmatch of DESI and \gaia from the MWS value-added catalogue, to obtain proper motion measurements.

\section{Different populations in the $L_Z$--\feh plane}\label{sec:id}
Now that we have discussed our simulations and the observational dataset, we will describe how we analysed the data from DESI to find this population of stars ejected from the GC. We also discuss stellar populations that might be misclassified as GC ejecta because of possible overlap in parameter space.

In the introduction we identified two characteristic properties that we expect for stars ejected from the GC over the past several billion years: we expect them to be metal rich and travel on initially radial trajectories. We discuss in Sect.~\ref{sec:simulations} how these stars oscillate in $L_X$ and $L_Y$, while $L_Z$ remains constant at 0. We therefore attempted to find a statistical overdensity of high-metallicity stars with an angular-momentum distribution centred at $L_Z=0$. Furthermore, we know the shape of the angular-momentum distribution in $L_Z$, since we expect this shape to be dominated by the observational uncertainties in the distances to individual stars, as discussed in Sect.~\ref{sec:simulations}.

\begin{figure}
    \centering
    \includegraphics[width=\linewidth]{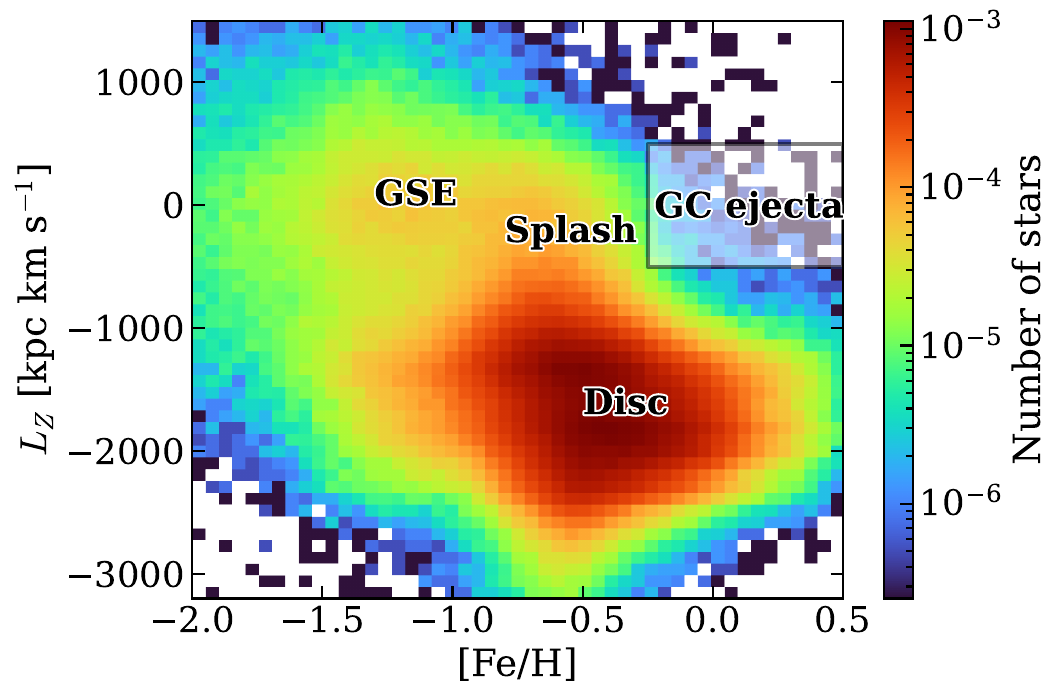}
    \caption{Histogram of \feh against $L_Z$ for stars in the DESI DR1 selection described in Sect.~\ref{sec:obs}. We highlight areas occupied by different galactic components and GC ejecta.}
    \label{fig:feh_lz}
\end{figure}

In order to identify an overdensity in \feh--$L_Z$ space, it is important to understand what other populations of stars overlap in this parameter space with GC ejecta. To gain intuition for this parameter space, we show metallicity against $L_Z$ for all stars in our observational dataset (see Sect.~\ref{sec:obs}) in Fig.~\ref{fig:feh_lz}.
At low metallicity ($-2<$ \feh$<-1$) and centred at $L_Z=0$ we can see the population known as Gaia-Sausage-Enceladus \citep[GSE;][]{Belokurov_2018, Helmi_2018, Haywood_2018}. GSE dominates the stellar halo in the solar neighbourhood and consists of metal-poor stars \citep[mean \feh$\sim-1
.2$;][]{Feuillet_2021} on highly eccentric orbits \citep[for a review, see][]{Deason_2024}. This population of stars is believed to be the remains of the most recent major merger of the Milky Way, which occurred $\sim10$ Gyr ago. The $L_Z$ distribution of GSE peaks roughly at $L_Z=0$, the same as the GC ejecta. However, unlike our expectation for GC ejecta, GSE does have a measurable width in $L_Z$ space by DESI. 

Metallicity is the other important differentiator: the metal-rich boundary of the GSE is typically found at around \feh $\sim-0.6$ \citep[e.g.][]{Myeong_2019, Hasselquist_2021, Horta_2023}, much lower in metallicity compared to typical stars expected from the GC \citep{Do_2015, Feldmeier_2017, Schodel_2020, Feldmeier-Krause_2025}. However, it is not excluded that there is a more metal-rich compact GSE remnant \citep[e.g.][]{Hasselquist_2021}. Finally, uncertainties in the measured \feh might lead to the misclassification of a GSE star.

The other population of stars that can end up in the high-metallicity, low-angular-momentum part of parameter space comes from the (thick) disc. In Fig.~\ref{fig:feh_lz} we can see that most stars reside at large negative $L_Z$, but a tail towards low angular momenta extends from this population. This tail is possibly related to the recently discovered population of relatively metal-rich (\feh$>-0.7$) stars on highly eccentric orbits known as the `Splash' \citep{Belokurov_2020}. These stars have little to no angular momentum and their relatively high metallicity means that they might overlap significantly with GC ejecta. A suggested interpretation of the Splash is that these stars were part of the proto-disc of the Galaxy and were kicked out during the merger event that created GSE. For detailed information on these and other halo populations, we refer to recent reviews \citep[e.g.][]{Rix_2013, Bland-Hawthorn_2016, Deason_2024, Hunt_2025}. \\

Now that we have a better understanding of the background stellar populations at high metallicity and low angular momentum, we will describe our model. We performed all our fits in $L_Z$ space in bins of metallicity. Furthermore, we separately treated dwarfs and giants, which we separated at $\log g =4$ as determined in the DESI MWS catalogue from RVS. We focussed on the region $|L_Z|<500$ \kpckms, since the population extending from large negative $L_Z$ can be described by a linear slope in this region. We refer to this population as the Splash. We described the GSE using a Gaussian centred at $L_Z=0$ and determined its width in $L_Z$ by performing an unbinned fit of this Gaussian and a linear slope to the distribution of $L_Z$. We limited this fit to the range $|L_Z| < 500$ \kpckms and $-1.7 < {\rm [Fe/H]} < -1.3$ since this parameter range is dominated by GSE stars in the DESI data. We find a standard deviation of $\sim 170$ \kpckms for giants and $\sim 410$ \kpckms for dwarfs and use these values to fix the width in $L_Z$ of GSE for our fits in metallicity bins. We left the slope of the Splash population as a free parameter.

Our probability density function to describe the distribution of stars in $L_Z$ within the range $[-L_{Z,\ \rm{max}}, L_{Z,\ \rm{max}}]$ per bin of metallicity is then
\begin{equation}\label{eq:model}
\begin{split}
    f(L_Z)\ =\ & p_{\rm GSE}\frac{1}{\sqrt{2\pi\sigma^2}}\exp\left(-\frac{L_Z^2}{2\sigma^2}\right)\frac{1}{Z_{\rm norm,\ GSE}}\\
    & + p_{\rm ej}\frac{1}{2b}\exp\left(\frac{-|L_Z|}{b}\right)\frac{1}{Z_{\rm norm,\ ej}}\\
    & + (1-p_{\rm GSE}-p_{\rm ej})\left(\frac{1}{2L_{Z,\ \rm{max}}}+aL_Z\right),
\end{split}
\end{equation}
where $p_{\rm GSE}$ and $p_{\rm ej}$ are the fractions of stars belonging to in the GSE and the GC ejecta, respectively, $\sigma$ is the width of the GSE in $L_Z$, $Z_{\rm norm,\ GSE}$ and $Z_{\rm norm,\ ej}$ are the integrals of the normal and Laplace distributions, respectively, over the range $[-L_{Z,\ \rm{max}}, L_{Z,\ \rm{max}}]$, $b$ is the scale parameter of the GC ejecta, and $a$ is the slope of the linear Splash population. For bins in metallicity, we performed unbinned fits to the probability density function for $L_{Z,\ \rm{max}}=500$ \kpckms using a Markov chain Monte Carlo approach. We used 100 walkers, with 500 burn-in steps and 1000 additional steps to explore the posterior and calculate percentile confidence intervals. The free parameters in this fit are $p_{\rm GSE}$, $p_{\rm ej}$, and $a$. We used uniform priors on $p_{\rm GSE}$ and $p_{\rm ej}$ between 0 and 1, and for $a$ we used a uniform prior for $|a|\leq1/(2L_{Z,\ \rm{max}})$, so that the posterior is positive for $|L_Z|<L_{Z,\ \rm{max}}$. Finally, we required that $1 - p_{\rm GSE} - p_{\rm ej} \geq 0$ so that our probability density function integrates to 1.

\section{Results}\label{sec:results}
\begin{figure*}
    \centering
    \begin{subfigure}{0.49\textwidth}
    \includegraphics[width=\linewidth]{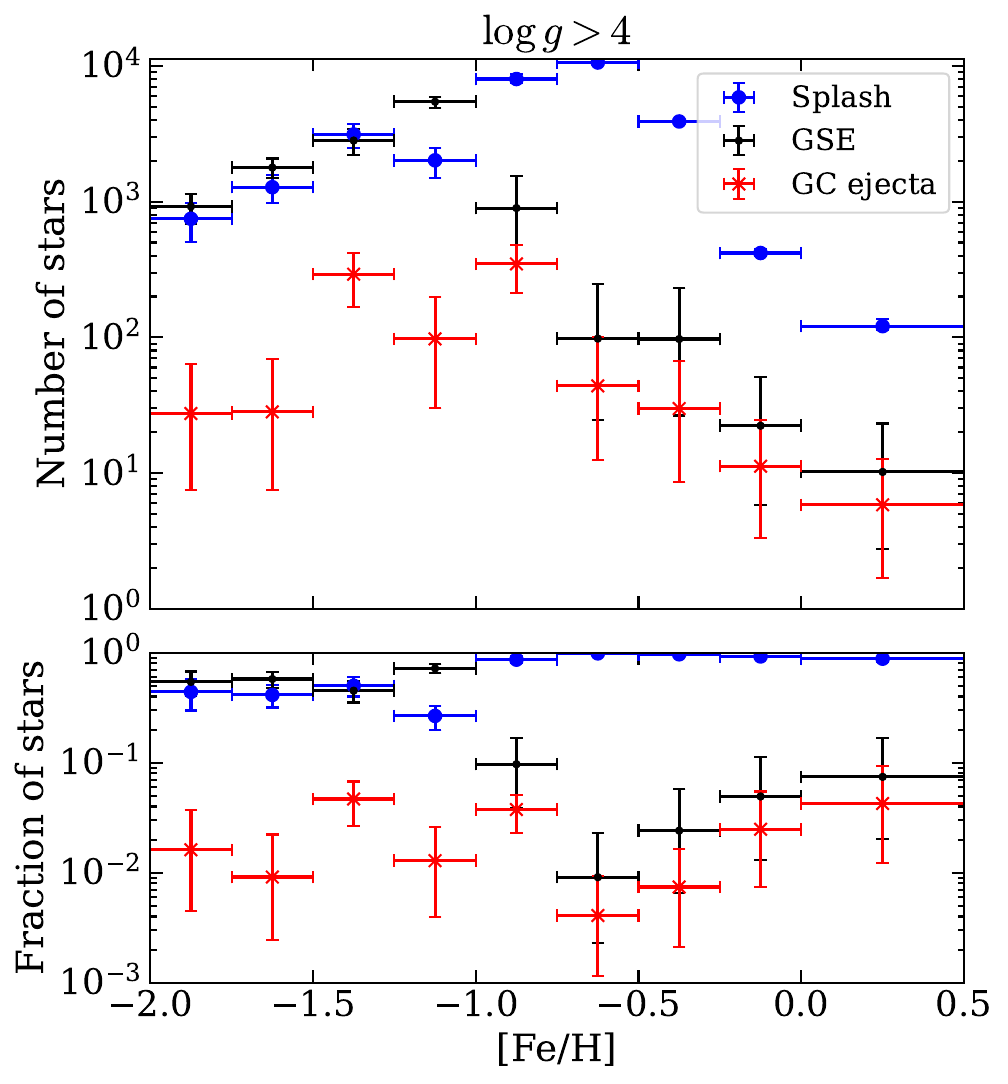}
    \end{subfigure}
    \begin{subfigure}{0.49\textwidth}
    \includegraphics[width=\linewidth]{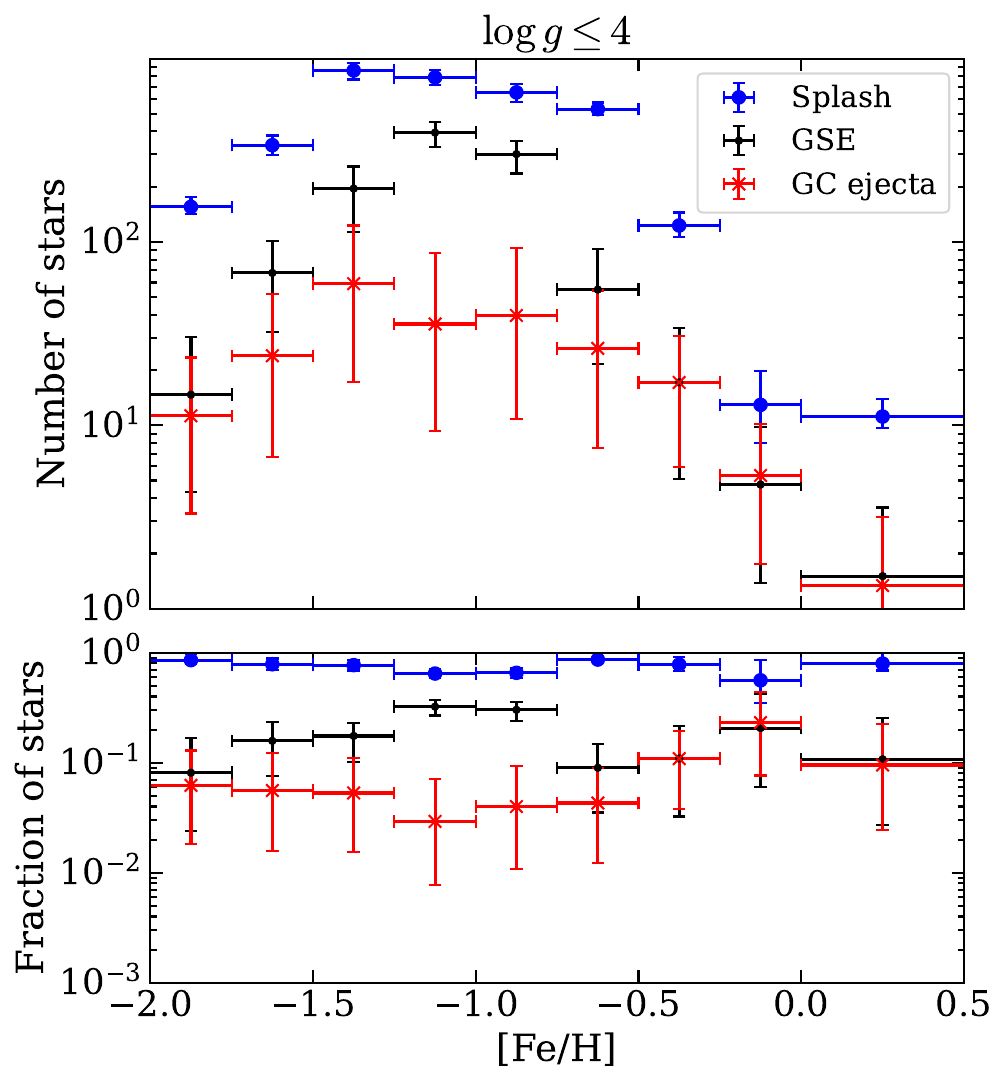}
    \end{subfigure}
    \caption{Relative contributions of Splash, GSE, and GC ejecta for stars with $|L_Z|<500$ \kpckms as a function of metallicity. The rows show the number of stars and fraction in each population, while the columns show dwarfs and giants. The horizontal error bars correspond to the bin size in \feh, and the vertical error bars are from the 16th and 84th percentiles of the posterior.}
    \label{fig:GSE_splash}
\end{figure*}

Now that we have discussed our model and the data, we will look at our results. In Fig.~\ref{fig:GSE_splash} we show the fraction and number of stars that belong to the Splash, GSE, and GC ejecta populations, as a function of metallicity.
The fraction of GC ejecta is always consistent with 0 if we consider that $p_{\rm ej}\geq0$ and the fact that the posteriors are highly non-Gaussian given this truncation. In fact, the posterior on $p_{\rm ej}$ usually peaks at $p_{\rm ej}=0$. For dwarfs we can see that at low metallicity the Splash and GSE contribute about equally to the stellar population, while for giants the Splash always dominates. Note that these numbers are not necessarily representative of the `true' fractions in the halo, since we did not correct for observational selection effects nor is our model optimised for detecting populations other than GC ejecta.

So far, we have not made any selections on angular momentum other than in $L_Z$. However, for GC ejecta, we expect that $L_X$ is also centred and concentrated around $L_X=0$. To improve our statistics, we performed an additional fit selecting only $L_X<300$ \kpckms to constrain the contribution from the GC ejecta population to the metallicity bin $-0.25<$ \feh $<0.5$. We chose this metallicity bin to encompass all GC ejecta. For both dwarfs and giants we again find a fraction of GC ejecta consistent with 0. The $95\%$ upper limit on the number of GC ejecta within this metallicity bin is about 38 dwarfs and 13 giants, for a total of about 51 stars. However, not all GC ejecta will be in our observational sample. The $95\%$ upper limit on the number of GC ejecta with $16<r<19$ and $|b|>40$ deg can be written as
\begin{equation}
    N_{\rm GC, ej}^{95}(16<r<19\ \wedge\ |b|>40\ {\rm deg}) = \frac{p_{\rm ej}^{95}N_{\rm [Fe/H]}}{C_{\rm spec}C_{\rm colour}C_{\rm Dec}C_{\rm L}},
\end{equation}
with $p_{\rm ej}^{95}$ the $95\%$ upper limit on the fraction of GC ejecta, $N_{\rm [Fe/H]}$ the number of stars in the metallicity bin, and $C_{\rm spec}$, $C_{\rm colour}$, $C_{\rm Dec}$, and $C_{\rm L}$ the completenesses on the spectroscopy, colour selection, Dec range, and angular momentum, respectively. The spectroscopic completeness of DESI DR1 is $20\%,$ as mentioned before, and from our simulations we find that the product of the other completeness factors is about $60\%$. The total number of GC ejected stars with $16<r<19$ at $|b|>40$ can therefore be no higher than $51/(0.2\cdot0.6)\simeq4.3\times10^2$ at the $95\%$ confidence limit, assuming that the ejection mechanism is isotropic.

The upper limit for the number of GC ejecta in DESI implies constraints on the ejection rate of stars from the GC over a timescale of billions of years, since GC ejecta will have accumulated in the halo for billions of years. Since we expect $\sim0.18*\frac{\eta}{10^{-5} {\rm yr^{-1}}}$ GC ejecta in DESI for an ejection rate $\eta$ (see Sect.~\ref{sec:sim_results}), the average ejection rate over this timescale has to be at most $\sim2.8\times10^{-3}$ yr$^{-1}$.

\section{Discussion}\label{sec:discussion}
In this section we provide context for our results, review some of our assumptions, and discuss future prospects. Since our simulations (excluding orbit integrations) are directly taken from the work of \citet{Verberne_2025}, we refer to that paper for a discussion of the ejection model and the progenitor binary population assumptions.

In this work we investigated whether we could identify an overdensity of high-metallicity stars at low angular momenta in the Galactic halo using data from DESI DR1. Such a population could point to Hills mechanism disruptions of binary stars near Sgr A* or past mergers of Sgr A* with intermediate-mass or massive black holes. Our analysis yields a non-detection that we used to constrain the GC ejection rate over a timescale of $\sim5$ Gyr, a novelty with respect to previous constraints obtained with unbound stars that were therefore limited to the shorter timescales of their fly times ($\sim100$ Myr). Specifically, we conclude that the average ejection rate over the past $\sim5$ Gyr cannot exceed $\sim2.8\times10^{-3}$ yr$^{-1}$. This rate is in line with previous estimates and constraints for the Hills mechanism, which tend to span the range $10^{-5}-10^{-3}$ yr$^{-1}$ \citep{Hills_1988, Yu_2003, Bromley_2012, Zhang_2013, Brown_2014, Evans_2022_I, Evans_2022, Marchetti_2022, Verberne_2024, Verberne_2025}. 

\subsection{Testing the ejection model dependence}\label{sec:ej_model}
Here we evaluate the dependence of our results on the specific ejection model assumed. A previous study showed that the velocity distribution of stars ejected from the GC is determined by the potential rather than the ejection velocity spectrum for stars travelling on bound orbits \citep{Rossi_2014}. We verified that this holds for our simulations by using different ejection velocity distributions. We find that the width of the zero-angular-momentum population and the number of expected GC ejecta in DESI are not significantly affected, which means that our results hold for any population of stars ejected isotropically from the GC and are thus ejection model independent. Anisotropic ejections could potentially bias our results if, for instance, stars are preferentially ejected in the Galactic plane, since DESI targets high latitude sources. Additionally, anisotropic ejections could change the angular-momentum distribution of GC ejecta, which would influence our results presented in Fig.~\ref{fig:GSE_splash}. 

\subsection{Testing the (non-)axisymmetric potential dependence}\label{sec:potential}
So far, we have used the axisymmetric \citet{McMillan_2017} potential for our orbit integration. The effect is that the intrinsic $L_Z$ distribution is a delta function at $L_Z=0$ and the spread in the observed $L_Z$ distribution is exclusively caused by observational uncertainties. However, the potential of the Galaxy contains non-axisymmetric components that will impart $L_Z$ onto GC ejecta. To investigate this effect, we made use of the potential defined in \citet[][excluding Sgr A*]{Hunter_2024}, which contains the nuclear star cluster, nuclear stellar disc, bar, thin and thick stellar discs, gas discs, dark halo, and spiral arms. Both the bar and spiral arms are non-axisymmetric and are rotating in this potential. We followed the same procedure we used for the \citet{McMillan_2017} potential and integrate the orbits using {\tt Agama}. The result of the non-axisymmetric components is that the $L_Z$ distribution becomes wider, with a scale parameter of $\sim1.5\times10^2$ \kpckms, compared to $\sim1.2\times10^2$ \kpckms for the \citet{McMillan_2017} potential. We also evaluated if the increased width in $L_Z$ affects our conclusions, by reanalysing the DESI data using our model (Eq.~\ref{eq:model}) with this increased width of the GC ejecta population. We find that the increased width of the $L_Z$ distribution does not significantly impact the ratios of stars we assigned to the Splash, GSE, or GC ejecta populations. Our rate and number constraints for GC ejecta increase by about $15\%$ if we assume the \citet{Hunter_2024} potential over the \citet{McMillan_2017} one. We therefore conclude that our conclusions are not significantly affected by the assumption that the potential is axisymmetric.

Separate from the adopted potential, a factor that might introduce additional angular momentum is scattering from individual stars. We implicitly  assumed in this work that this is negligible, but especially for stars that experience multiple pericentre passages through the dense GC region, scattering might become important. Since we only considered stars with apocentres outside the GC, we effectively limited the number of orbits for these stars. Indeed, we find that for stars with an apocentre greater than 1 kpc, the number of orbits is at most a few tens in our simulations. We therefore assumed that scattering by individual stars is not important.

\section{Summary and conclusion}\label{sec:summary}
In this work we searched for the population of GC ejecta that we expect has accumulated in the stellar halo over billions of years via the Hills mechanism and massive black hole binary ejections. To this purpose, we used the newly released DESI DR1 in combination with simulations to predict the angular-momentum distribution of stars ejected from the GC. We find that the radial distribution of GC ejecta between 1 and 100 kpc from the GC can be described by a single power law with $\rho\propto r^{-3.41\pm0.01}$. Gravitational focussing by the disc means that many stars ejected from the GC will have ended up in the stellar disc, making identification more challenging due to the large number of disc stars. We built a mixture model to describe the $L_Z$ distribution around $L_Z=0$ as a function of metallicity and fitted this to the DESI data. We find that on a statistical level there is no overdensity at high metallicities of stars on low $L_Z$ orbits, as would be expected for GC ejecta. Based on this null-detection, we place an ejection-model-independent upper limit on the ejection rate of stars from the GC of $\sim2.8\times10^{-3}$ yr$^{-1}$. We also place a 95\% upper limit on the number of GC ejecta at $|b|>40$ deg and $16<r<19$ of $4.3\times10^2$ stars. 

The analysis presented here can be repeated for other large spectroscopic surveys, for instance using the \gaia DR3 radial velocity subsample \citep{Gaia_2023}, {LAMOST} \citep{Cui_2012}, or {APOGEE} \citep{Majewski_2017}. Furthermore, the upcoming \gaia DR4 and DESI DR2 will significantly increase the number of sources to which this analysis can be applied. DESI DR2, for instance, is expected to contain about three times more stars than DR1 \citep{Koposov_2025}. Our method would be enhanced by considering additional tracers, such as elemental abundances, which would add weight to any claimed GC ejecta discoveries \citep[see e.g.][]{Hattori_2025}.

Our results can be used to constrain ejection mechanisms; a particularly interesting application is constraining the merger history of Sgr A*. Gravitational slingshots of single stars by a massive binary black hole eject stars with properties similar to the Hills mechanism \citep{Yu_2003}. \citet{Evans_2023} already used this in combination with the lack of uncontroversial GC ejecta in \gaia DR3 to determine that mergers of Sgr A* with a possible companion of $>500$ M$_\odot$ cannot have happened within the past 10 Myr. Expanding on this, we believe that our non-detection of a population of bound ejecta constrains the merger history of Sgr A* to a timescale of billions of years, something we know very little about.

\begin{acknowledgements}
The authors thank Koen Kuijken, Manuel Cavieres Carrera, Adrian Price-Whelan, Carrie Filion, and  Danny Horta for useful discussions.
EMR and ZP acknowledge support from the European Research Council (ERC) grant number: 101002511/project acronym: VEGA\_P. SK acknowledges support from the Science \& Technology Facilities Council (STFC) grant ST/Y001001/1. We acknowledge the Gaia Project Scientist Support Team and the Gaia Data Processing and Analysis Consortium (DPAC). This work has made use of data from the European Space Agency (ESA) mission {\it Gaia} (\url{https://www.cosmos.esa.int/gaia}), processed by the {\it Gaia} Data Processing and Analysis Consortium (DPAC, \url{https://www.cosmos.esa.int/web/gaia/dpac/consortium}). Funding for the DPAC has been provided by national institutions, in particular the institutions participating in the {\it Gaia} Multilateral Agreement. This research used data obtained with the Dark Energy Spectroscopic Instrument (DESI). DESI construction and operations is managed by the Lawrence Berkeley National Laboratory. This material is based upon work supported by the U.S. Department of Energy, Office of Science, Office of High-Energy Physics, under Contract No. DE–AC02–05CH11231, and by the National Energy Research Scientific Computing Center, a DOE Office of Science User Facility under the same contract. Additional support for DESI was provided by the U.S. National Science Foundation (NSF), Division of Astronomical Sciences under Contract No. AST-0950945 to the NSF’s National Optical-Infrared Astronomy Research Laboratory; the Science and Technology Facilities Council of the United Kingdom; the Gordon and Betty Moore Foundation; the Heising-Simons Foundation; the French Alternative Energies and Atomic Energy Commission (CEA); the National Council of Humanities, Science and Technology of Mexico (CONAHCYT); the Ministry of Science and Innovation of Spain (MICINN), and by the DESI Member Institutions: www.desi.lbl.gov/collaborating-institutions. The DESI collaboration is honored to be permitted to conduct scientific research on I’oligam Du’ag (Kitt Peak), a mountain with particular significance to the Tohono O’odham Nation. Any opinions, findings, and conclusions or recommendations expressed in this material are those of the author(s) and do not necessarily reflect the views of the U.S. National Science Foundation, the U.S. Department of Energy, or any of the listed funding agencies.
For the purpose of open access, the author has applied a Creative
Commons Attribution (CC BY) licence to any Author Accepted Manuscript version arising from this submission.\\

Software: \texttt{NumPy} \citep{Harris_2020}, \texttt{SciPy} \citep{2020SciPy-NMeth}, \texttt{Matplotlib} \citep{Hunter_2007}, \texttt{Astropy} \citep{astropy:2013, astropy:2018, astropy:2022}, \texttt{isochrones} \citep{Isochrones_2015}, \texttt{Speedystar} \citep{Contigiani_2019, Evans_2022_I}, \texttt{emcee} \citep{Foreman_2013}, {\tt Agama} \citep{Vasiliev_2019}.
\end{acknowledgements}

\bibliographystyle{aa}
\bibliography{mybib}

\end{document}